\documentclass[prd,amsmath,notitlepage,eqsecnum,twocolumn,superscriptaddress,nofootinbib]{revtex4-2}

\pdfoutput=1
\usepackage[utf8]{inputenc}

\usepackage{graphicx}
\usepackage{float}
\usepackage{epstopdf,cancel}
\usepackage{epsf,latexsym,bbm,euscript}
\usepackage{amssymb,amsmath,bm}
\usepackage{mathtools}
\usepackage{times,graphics}
\usepackage{soul,xcolor}
\usepackage{mathtools}
\usepackage{tensor}
\usepackage[normalem]{ulem}
\usepackage{setspace}
\usepackage{enumitem}
\usepackage{accents}
\usepackage{booktabs}
\usepackage{microtype}
\usepackage[caption=false]{subfig}
\usepackage{etoolbox}
\usepackage{titlesec}
\usepackage{siunitx}

\makeatletter
\newcommand*{\defeq}{\mathrel{\rlap{%
			\raisebox{0.3ex}{$\m@th\cdot$}}%
		\raisebox{-0.3ex}{$\m@th\cdot$}}%
	=}
\newcommand*{\eqdef}{=\mathrel{\rlap{%
			\raisebox{0.3ex}{$\m@th\cdot$}}%
		\raisebox{-0.3ex}{$\m@th\cdot$}}%
}
\makeatother

\makeatletter
\g@addto@macro\bfseries{\boldmath}
\makeatother

\makeatletter
\def\thesubsection{\thesection.\arabic{subsection}}
\def\p@subsection{}
\makeatother

\titleformat{\section}[hang]
{\normalfont\normalsize\bfseries\MakeUppercase}
{\thesection.}{0.75em}
{\raggedright}

\titleformat{\subsection}[hang]
{\normalfont\normalsize\bfseries}
{\hspace{1em}\thesubsection.}{0.5em}
{\hspace{0pt}\raggedright}

\titleformat{\subsubsection}[hang]
{\normalfont\normalsize\bfseries}
{\thesubsubsection.}{1em}{\raggedright}

\titlespacing*{\section}{0pt}{4mm}{1mm}
\titlespacing*{\subsection}{0pt}{3mm}{0.5mm}
\titlespacing*{\subsubsection}{0pt}{2.25mm}{0.25mm}

\apptocmd{\appendix}{%
	\titleformat{\section}[hang]
	{\normalfont\normalsize\bfseries}
	{Appendix \thesection:}{0.5em}{\raggedright}%
	\titlespacing*{\section}{0pt}{3mm}{0.5mm}%
}{}{}

\usepackage{scalerel}
\usepackage{tikz}
\usetikzlibrary{svg.path}
\definecolor{orcidlogocol}{HTML}{A6CE39}
\tikzset{
	orcidlogo/.pic={
		\fill[orcidlogocol] svg{M256,128c0,70.7-57.3,128-128,128C57.3,256,0,198.7,0,128C0,57.3,57.3,0,128,0C198.7,0,256,57.3,256,128z};
		\fill[white] svg{M86.3,186.2H70.9V79.1h15.4v48.4V186.2z}
		svg{M108.9,79.1h41.6c39.6,0,57,28.3,57,53.6c0,27.5-21.5,53.6-56.8,53.6h-41.8V79.1z M124.3,172.4h24.5c34.9,0,42.9-26.5,42.9-39.7c0-21.5-13.7-39.7-43.7-39.7h-23.7V172.4z}
		svg{M88.7,56.8c0,5.5-4.5,10.1-10.1,10.1c-5.6,0-10.1-4.6-10.1-10.1c0-5.6,4.5-10.1,10.1-10.1C84.2,46.7,88.7,51.3,88.7,56.8z};
	}
}
\newcommand\orcidlink[1]{\href{https://orcid.org/#1}{\mbox{\scalerel*{
				\begin{tikzpicture}[yscale=-1,transform shape]
					\pic{orcidlogo};
			\end{tikzpicture}}{X}}}}

\newcommand{\g}{{\textsl{g}}}

\newcommand{\be}{\begin{equation}}
	\newcommand{\ee}{\end{equation}}
\newcommand{\ba}{\begin{eqnarray}}
	\newcommand{\ea}{\end{eqnarray}}
\def\half{\tfrac{1}{2}}

\def\6{{\langle}}
\def\9{{\rangle}}

\def\deo{{\Delta_1}}
\def\dt{{\Delta_2}}
\def\xfm{{x^{(\mu)}}}

\def\ga{\mathfrak{a}}

\def\ak{\mathsf{k}}
\def\al{\mathsf{l}}
\def\an{\mathsf{n}}
\def\am{\mathsf{m}}
\def\ave{\mathsf{e}}
\def\aw{\mathsf{w}}

\usepackage{fancyhdr}
\definecolor{arxivlink}{RGB}{0,140,220}
\definecolor{doilink}{RGB}{10,30,160}

\usepackage{url,hyperref}
\hypersetup{colorlinks,linkcolor={blue!55!black},citecolor={red!50!black},urlcolor={doilink},breaklinks=true}

\begin{document}
	
	\title{Covariant variation for point-particle Lagrangians}
	
	\author{Finnian Gray\orcidlink{0000-0003-0420-4046}}
	\email{finnian.gray@mq.edu.au}
	\affiliation{School of Mathematical and Physical Sciences, Macquarie University, NSW 2109, Australia}
	
	\author{Sebastian Murk\orcidlink{0000-0001-7296-0420}}
	\email{sebastian.murk@matfyz.cuni.cz}
	\affiliation{Faculty of Mathematics and Physics, Charles University, Ke Karlovu 3, 121 16 Praha 2, Czech Republic}
	
	\author{Daniel R.\ Terno\orcidlink{0000-0002-0779-0100}}
	\email{daniel.terno@mq.edu.au}
	\affiliation{School of Mathematical and Physical Sciences, Macquarie University, NSW 2109, Australia}
	
	\begin{abstract}
		Structureless test particles in general relativity follow geodesics.
		For extended bodies, higher-order multipole moments lead to departures from geodesic motion; in particular, spinning test bodies obey the Mathisson--Papapetrou--Dixon (MPD) equations. Similarly, the leading correction to the eikonal approximation for electromagnetic-wave propagation can be formulated as the nongeodesic propagation of spinning null particles. When the resulting equations are treated as standalone worldline models, with the relevant dynamical quantities defined only along the representative worldline, their variational formulation requires particular care.
		Following DeWitt's construction, we distinguish several types of variation, define the corresponding covariant variation for each, and identify the role of parallel transport in models that couple worldline variables to tensor fields.
		This framework simplifies the variational treatment of the MPD equations and yields a simple Lagrangian for the null-particle model of light propagation.
	\end{abstract}
	
	\maketitle
	\thispagestyle{fancy}
	
	\section{Introduction}
	Astrophysical observations rely primarily on the electromagnetic and gravitational waves that reach us from distant sources.
	The study of their propagation is often technically difficult, and therefore it is commonly performed within the eikonal (geometric optics) approximation.
	This approach is based on the asymptotic expansion of wave solutions in inverse powers of the characteristic frequency $\omega$ and retains only the leading-order terms~\cite{Sachs:ProcRSocA:1961,Kristian.Sachs:AstrophysJ:1966,Ehlers:ZNaturforschA:1967,Seitz.Schneider.Ehlers:CQG:1994,Frolov.Shoom:PRD:2011}.
	The resulting short-wave limit provides the standard description of null propagation of fictitious massless point particles in curved spacetime. It underlies two classical tests of general relativity~\cite{Will:Book:2018} and describes polarization rotation in gravitational fields~\cite{Oancea.etal:PRD:2020,Andersson.etal:PRD:2021,Frolov.Shoom:JCAP:2024}.
	
	For more than a decade, increasing attention has focused on corrections beyond geometric optics, in particular on the next order in the $1/\omega$ expansion, at which polarization affects the trajectory itself.
	The corresponding equations describe null trajectories that are no longer geodesic and can be interpreted as the motion of massless particles subject to polarization-dependent forces~\cite{Dolan:1801.02273,Dolan:IJMPD:2018,Oancea.etal:PRD:2020,Frolov:PRD:2020,Andersson.etal:PRD:2021,Andersson.Oancea:CQG:2023}.
	With suitable truncations, these equations can be reproduced from various model point-particle Lagrangians.
	In the massless case, these effective descriptions can be compared with the many models of massless spinning particles proposed in the literature~\cite{Souriau:AnnInstHenriPoincareA:1974,Mashhoon:AnnPhys:1975,Saturnini:Thesis:1976,Bailyn.Ragusa:PRD:1977,Bailyn.Ragusa:PRD:1981,Semerak:PRD:2015,Deriglazov.Ramirez:IJMPD:2017}.
	By contrast, massive spinning bodies are conventionally described by the Mathisson--Papapetrou--Dixon (MPD) equations, which are obtained from a multipole expansion of the stress-energy tensor of an extended body~\cite{Mathisson:ActaPhysPol:1937,Papapetrou:ProcRoySocLondA:1951,Dixon:ProcRoySocLondA:1970,Dixon:PhilTransRoySocLondA:1974}.
	
	Such effective Lagrangian formulations are useful not only because they reproduce the equations of motion, but also because they provide a systematic framework for deriving conserved quantities and relating them to explicit and hidden symmetries through Noether-type arguments.
	However, exploiting this framework requires a careful treatment of variations. Since the relevant dynamical variables are defined on curved spacetime and involve constrained internal degrees of freedom, calculations quickly become cumbersome and often obscure covariance.
	As a result, the variational structure in the literature is not always presented in a fully transparent or consistent form.
	
	A covariant variational calculus provides a systematic way to address these difficulties.
	In DeWitt's covariant-variation formalism \cite{DeWitt:Book:2011}, tensorial quantities at nearby spacetime points are compared by parallel transport such that the variation can be organized in terms of covariant objects from the outset.
	Its application to general relativistic binary mechanics was reviewed by Steinhoff~\cite{Steinhoff:Chapter:2015}.
	These ideas are closely related to the notion of horizontal and vertical covariant derivatives in configuration or phase space~\cite{Sharafutdinov:Book:1994,Cariglia.etal:PRD:2013,Andersson.etal:2507.11965}.
	
	Our goal is twofold: first, we present the variational procedure in a manifestly covariant form and clarify the roles of the different notions of variation; second, we apply this framework to effective Lagrangians relevant for polarization-corrected propagation and show how it enables a clean and systematic derivation of the equations of motion.
	We first illustrate our method on model Lagrangians of the form
	\begin{equation}\label{modeL}
		L=L\big(x^\mu(\tau), \dot x^\mu(\tau)\big),
	\end{equation}
	where $x^\mu(\tau)$ denotes a worldline parametrized by $\tau$, $\dot{x}^\mu(\tau)$ is its tangent vector, and $\tensor{\g}{_\mu_\nu}(x)$ is a fixed background metric.
	We then extend the formalism to worldline models with internal degrees of freedom and give simple derivations of the MPD equations and the associated conservation laws.
	Finally, we present the Lagrangian that generates the equations of motion for massless particles whose nongeodesic null trajectories describe light rays in the post-eikonal approximation.
	
	It is important to distinguish between two types of tensorial objects to which the variational techniques are applied.
	The metric $\tensor{\g}{_\mu_\nu}(x)$ and spacetime fields $\phi$ are defined on the spacetime manifold $\mathcal{M}$, or on an open coordinate patch, and are sections of the corresponding tensor bundles over $\mathcal{M}$.
	By contrast, the tangent vector $\al=l^\mu\partial_\mu$, with $l^\mu\defeq \dot{x}^\mu$, and other worldline tensors $\psi$ (e.g., the vectors of a tetrad propagated with the particle) are defined only along the worldline.
	More precisely, such a tetrad is a section of the frame bundle pulled back along the worldline map $\tau\mapsto x^\mu(\tau)$, with the appropriate normalization or orthonormality constraints imposed.
	Its individual vectors are sections of the pulled-back tangent bundle $x^*T\mathcal{M}$.
	Similarly, a point-particle Lagrangian $L(x,\dot{x})$ is a scalar function on $T\mathcal{M}$ and becomes a scalar function along the worldline once $x^\mu(\tau)$ is specified.

	The remainder of this article is organized as follows.
	In Sec.~\ref{sec:covariant.variation}, we introduce the different notions of variation and explain their mutual relations.
	In Sec.~\ref{S:app}, we first apply the formalism to spinless point particles and then to massive spinning particles, deriving the MPD equations and the associated conservation laws.
	In Sec.~\ref{sec:Lagrangian.mechanics}, we present the Lagrangian that yields the first-order post-eikonal equations in the gauge adopted in Ref.~\cite{Frolov:PRD:2020}.
	Finally, in Sec.~\ref{sec:discussion.and.conclusions}, we discuss the physical implications of our results and outline avenues for future work.
	
	Throughout the article, we denote generic tensorial objects by $\Phi$.
	When the distinction is relevant, we write $\phi=\phi(x)$ for spacetime fields and $\psi=\psi\big(x(\tau),\tau\big)$ for tensors defined along the worldline.
	A trajectory is a map $\Sigma \ni \tau\mapsto x(\tau)\in \mathcal{M}$, where $\Sigma\subseteq\mathbb{R}$ is the domain of the evolution parameter and $\mathcal{M}$ is the spacetime manifold.
	The tangent to the trajectory $\al(\tau)$ has coordinate components $l^\mu=dx^\mu/d\tau\equiv \dot{x}^\mu$.
	Index-free four-vectors are denoted by the {\sf Sans Serif} font, e.g., $\al$, $\ave$, $\aw$, and by the Computer Modern font (the default font family in \LaTeX) otherwise, e.g., $l^\mu$, $e^\mu$, $w^\mu$.
	The labels for vectors in an orthonormal tetrad are enclosed by parentheses whenever any of their co- and/or contravariant components are referred to explicitly [cf.\ Eq.~\eqref{eq:master}].
	If no components are referenced explicitly we omit the parentheses to reduce notational clutter [cf.\ Eq.~\eqref{eq:m.mbar}].
	We use the metric signature $(-,+,+,+)$ and define the Riemann tensor by $[\nabla_{\mu},\nabla_\nu]V^\rho\defeq R^\rho{}_{\sigma\mu\nu}V^\sigma$.

	\section{Covariant variation} \label{sec:covariant.variation}
	We follow DeWitt's formulation of covariant variation~\cite{DeWitt:Book:2011,Steinhoff:Chapter:2015}.
	Our contribution is to distinguish explicitly between the different types of variation needed for worldline variational problems and to construct their corresponding covariant extensions.
	In Sec.~\ref{S:app}, we show how these distinctions lead to simple derivations of the associated conservation laws.
	
	We first consider the variation of a tensorial object $\Phi$ induced by an infinitesimal diffeomorphism generated by a vector field $\bm{\xi}$,
	\begin{equation}
		\bar{x}^\mu = x^\mu + \epsilon\,\xi^\mu(x) + \mathcal{O}(\epsilon^2) .
		\label{eq:infinitesimal.diffeomorphism}
	\end{equation}
	In what follows, all infinitesimal equalities are understood in this sense, and we omit $\mathcal{O}(\epsilon^2)$ terms to reduce notational clutter.
	
	For any tensorial object $\Phi$ transforming under a representation $D$ of $\mathrm{GL}(4)$, with $J\equiv\partial\bar{x}/\partial x$ denoting the corresponding Jacobian, the diffeomorphism acts as
	\begin{equation}
		\bar{\Phi}(\bar{x}) = D(J)\,\Phi(x).
	\end{equation}
	Expanding to first order in $\epsilon$ gives
	\begin{equation}
		\bar{\Phi}(x + \epsilon\,\xi) = (I + \epsilon\,G^\mu_\nu\,\partial_\mu \xi^\nu)\,\Phi(x),
	\end{equation}
	where $I$ denotes the identity operator in the representation space of $\Phi$ and $G^\mu_\nu$ are the generators of $D$.
	
	The action of $G$ on arbitrary tensors is built from the primitives
	\begin{equation}\label{eq: Rep vecs}
		(G^\mu_\nu\Phi)^\alpha = \delta^\alpha_\nu \Phi^\mu, \qquad
		(G^\mu_\nu\Phi)_\beta=-\delta^\mu_\beta\Phi_\nu.
	\end{equation}
	For a scalar density of weight $w$, transforming as $\bar{\Phi}=(\det J)^{-w}\,\Phi$, one has
	\begin{equation}
		G^\mu_\nu\Phi=-w\, \delta^\mu_\nu \Phi.
	\end{equation}
	Successive transformations lead to the commutation relation \cite{Steinhoff:Chapter:2015}
	\begin{equation}
		[G^\mu_\nu, G^\alpha_\beta]\Phi = \big(\delta^\mu_\beta G^\alpha_\nu - \delta^\alpha_\nu G^\mu_\beta\big) \Phi.
		\label{eq: Rep Commutator}
	\end{equation}
	
	Several types of variations are needed to formulate variational principles covariantly.
	We consider the following four elementary variations: the same-point variation $\delta$, the shift variation $\delta_1$, the total variation $\delta_2$, and the parallel-transport variation $\delta_\|$.
	
	The same-point variation (also referred to as modified variation) is defined only for fields as
	\begin{equation}
		\delta\phi \defeq \bar{\phi}(\bar{x}) - \phi(\bar{x}).
		\label{def:same-point.variation}
	\end{equation}
	It is a geometric quantity with the same transformation law as $\phi$.
These variations act on the tensorial variables defined at $x(\tau)$, but do not move the worldline point itself.
	
	The shift variation $\delta_1\phi$ is also defined only for fields,
	\begin{equation}
		\delta_1 \phi \defeq \phi(\bar{x}) - \phi(x) = \epsilon\, \xi^\mu \partial_\mu \phi.
		\label{def:shift.variation}
	\end{equation}
	When a spacetime field is pulled back to the worldline, we extend $\delta_1$ to its derivative along the parameter $\tau$ by requiring it to commute with $d/d\tau$:
	\begin{align}
		\delta_1 \dot{\phi} \defeq \frac{d}{d\tau} \delta_1 \phi = \epsilon\, \frac{d}{d\tau} \big(\xi^\mu \partial_\mu \phi\big) .
	\end{align}
	
	On the other hand, the total variation
	\begin{equation}
		\delta_2\Phi \defeq \bar{\Phi}(\bar{x}) - \Phi(x) = \epsilon\, G^\mu_\nu\, \xi^\nu_{,\mu} \Phi
		\label{def:total.variation}
	\end{equation}
	is defined for both fields and worldline tensors. Finally, to compare tensors at $x$ and $\bar{x}$, we define the parallel-transported value of $\Phi(x)$ at $\bar{x}$ by
	\begin{align}
		\begin{aligned}
			\Phi_\|(\bar{x}) & \defeq \Phi(x) - \epsilon\, \Gamma^\nu_{\mu\rho}(x)\, \xi^\rho(x)\, G^\mu_\nu\, \Phi(x) \\
			&\equiv \Phi(x)+\delta_\|\Phi,
		\end{aligned}
		\label{def:parallel.transport.variation}
	\end{align}
	where $\Gamma^\mu_{\nu\rho}$ are the Christoffel symbols of the Levi-Civita connection.
	
	For fields, the same-point variation is the difference between the total and shift variations,
	\begin{equation}
		\delta\phi=\delta_2\phi-\delta_1\phi=-\epsilon \pounds_\xi\phi,
		\label{delta-r}
	\end{equation}
	with the Lie derivative given by	
	\begin{equation}
		\pounds_\xi\phi = \big(\xi^\mu \partial_\mu - G^\mu_\nu\, \xi^\nu_{,\mu}\big) \phi.
		\label{Lie}
	\end{equation}
	
	Using parallel transport, one can introduce covariant variations~\cite{DeWitt:Book:2011}. Since these are obtained by comparing the elementary variations $\delta_1$ and $\delta_2$ with the parallel-transport variation $\delta_\|$, we denote them by $\Delta_1$ and $\Delta_2$, respectively. The variation
	\begin{equation}
		\Delta_1\phi \defeq \phi(\bar{x}) - \phi_\|(\bar{x})
		\label{def:Delta1}
	\end{equation}
	is defined for fields, while
	\begin{equation}
		\Delta_2\Phi \defeq \bar{\Phi}(\bar{x}) - \Phi_\|(\bar{x})
		\label{def:Delta2}
	\end{equation}
	is defined for any tensorial object.
	
	The covariant variation $\Delta_1$ is given explicitly as
	\begin{equation}
		\Delta_1\phi = \delta_1\phi - \delta_\|\phi = \epsilon\, \xi^\mu \nabla_\mu \phi,
		\label{eq:Delta1.explicit}
	\end{equation}
	where $\nabla$ denotes the covariant derivative. Similarly,
	\begin{equation}
		\Delta_2\Phi = \delta_2\Phi - \delta_\|\Phi = \epsilon \big(\nabla_\mu\xi^\nu\big) G^\mu_\nu\, \Phi
		\label{eq:Delta2.explicit}
	\end{equation}
	for generic tensorial objects. Consequently, for fields,
	\begin{equation}
		\delta\phi = - \epsilon \pounds_\xi \phi = (\Delta_2 - \Delta_1) \phi.
	\end{equation}
	Since $\nabla\g=0$, the metric is inert under $\Delta_1$,
	\begin{equation}
		\Delta_1 \g_{\mu\nu} \equiv 0.
		\label{d1g}
	\end{equation}
	However, it is inert under the action of $\Delta_2$ if and only if $\xi$ is a Killing vector field, since
	\begin{equation}
		\Delta_2 \g_{\mu\nu} = -\epsilon\,(\xi_{\mu;\nu}+\xi_{\nu;\mu})=-\epsilon\pounds_\xi \g_{\mu\nu}.
		\label{d2g}
	\end{equation}
	Consequently, for any vector $\Phi^\mu$ to which the corresponding variation applies,
	\begin{equation}
		\Delta_i \Phi_\nu = \Delta_i (\g_{\mu\nu}\Phi^\mu) = (\Delta_i \g_{\mu\nu}) \Phi^\mu + \g_{\mu\nu} \Delta_i \Phi^\mu,
		\;\; i=1,2.
	\end{equation}
	Thus $\Delta_1$ commutes with index manipulations, while $\Delta_2$ does so only when $\xi$ is Killing.
	
	The variations $\delta_1$ and $\Delta_1$ are not intrinsically defined for general worldline tensors. The tangent vector is exceptional in the sense that its components can be obtained locally as derivatives of the coordinate functions $\xfm(x) \!=\! x^\mu$ evaluated along the worldline $x(\tau)$, with the parenthesized index understood as a coordinate label rather than a tensor index. This allows the definition of $\Delta_1$ to be extended consistently to $l^\mu \!=\! d\xfm/d\tau \!=\! \nabla_\tau x^{(\mu)}$.
	
	We first note that $\xfm(x)$ satisfies
	\begin{equation}
		\delta_2\xfm=0, \qquad \delta\xfm=-\delta_1\xfm=-\epsilon\,\xi^\mu,
	\end{equation}
	and thus
	\begin{equation}
		\dt\xfm=0, \quad \deo\xfm=-\delta\xfm=\epsilon\pounds_\xi\xfm=\epsilon\,\xi^\mu.
	\end{equation}
	Second, the covariant worldline derivative is given by
	\begin{equation}
		\nabla_\tau\Phi = \dot{\Phi}[x(\tau)] + \Gamma^\mu_{\nu\rho}\, l^\rho\, G^\nu_\mu\, \Phi.
	\end{equation}
	When a spacetime field $\phi(x)$ is pulled back to the worldline, $\dot{\phi} \equiv l^\mu \phi_{,\mu}$, and therefore $\nabla_\tau\phi=l^\mu\phi_{;\mu}$.
	
	Under the infinitesimal diffeomorphism \eqref{eq:infinitesimal.diffeomorphism}, the tangent vector is pushed forward from $x$ to $\bar{x}$, yielding
	\begin{align}
		\begin{aligned}
			\delta_2 \dot{x}^{(\mu)} &= \bar l^\mu(\bar{x}) - l^\mu(x) \\
			&= l^\mu(x) + \epsilon\, \xi^\mu_{,\nu}(x) l^\nu(x) - l^\mu(x) = \epsilon\, l^\nu \xi^\mu_{,\nu}.
		\end{aligned}
	\end{align}
	Thus, for the coordinate functions, taking the total variation of their $\tau$ derivative is equivalent to differentiating their shift variation,
	\begin{align}
		\delta_2 \frac{d}{d\tau} x^{(\mu)} = \frac{d}{d\tau} \delta_1 x^{(\mu)} .
	\end{align}
	The corresponding covariant variation of the tangent vector is
	\begin{align}
		\begin{aligned}
			\dt l^\mu &= \delta_2 l^\mu - \delta_\| l^\mu \\
			&= \epsilon\,l^\nu\xi^\mu_{,\nu} + \epsilon\,\Gamma^\mu_{\nu\rho}l^\nu\xi^\rho = \epsilon\,\xi^\mu_{;\nu}l^\nu .
		\end{aligned}
	\end{align}
	Equivalently,
	\begin{align}
		\dt l^\mu \equiv \dt\nabla_\tau x^{(\mu)} = \nabla_\tau\deo x^{(\mu)} = \epsilon\,\xi^\mu_{;\nu}l^\nu .
	\end{align}
	
	On the other hand, we can extend $\delta_1$ to the tangent vector by requiring it to commute with $d/d\tau$ on the coordinate scalar functions,
	\begin{equation}
		\delta_1 \dot{x}^{(\mu)} \defeq \frac{d}{d\tau} \delta_1 x^{(\mu)} = \frac{d(\bar{x}^\mu-x^\mu)}{d\tau} = \epsilon\,\dot{\xi}^\mu \equiv \delta_2\dot{x}^{(\mu)} .
	\end{equation}
	Equivalently, the variation of the tangent is defined as the tangent to the varied trajectory,
	\begin{equation}
		\delta_1 \dot{x}^{(\mu)} = \bar{l}^\mu - l^\mu = \frac{d}{d\tau} \delta_1 x^{(\mu)} \equiv \delta_2\dot{x}^{(\mu)} .
	\end{equation}
	This extension satisfies
	\begin{equation}\label{eq: D_1 l}
		\delta_1 \dot{x}^{(\mu)} = \delta_2 \dot{x}^{(\mu)}, \qquad
		(\deo l)^\mu = (\dt l)^\mu = \epsilon\, \xi^\mu_{;\nu} l^\nu.
	\end{equation}
	However, this equality is not preserved under lowering the index. Indeed,
	\begin{equation}
		\Delta_1 l_\mu \defeq \Delta_1(\g_{\mu\nu} l^\nu) = \epsilon\, \g_{\mu\nu} \xi^\nu_{;\rho} l^\rho = \epsilon\, l^\nu \xi_{\mu;\nu} \neq \Delta_2 l_\mu,
		\label{l-mis}
	\end{equation}
	whereas
	\begin{equation}
		\Delta_2 l_\mu = -\epsilon\, l^\nu \xi_{\nu;\mu}.
	\end{equation}
	Therefore
	\begin{equation}
		(\Delta_1-\Delta_2) l_\mu = 2\epsilon\, l^\nu \xi_{(\mu;\nu)}.
	\end{equation}
	By contrast, with the above extension of $\delta_1$ to the tangent vector, the same-point variation can be extended to $l^\mu$ by $\delta l^\mu\defeq\delta_2 l^\mu-\delta_1 l^\mu$. Since $\delta_1 l^\mu=\delta_2 l^\mu$, we have
	\begin{align}
		(\Delta_1-\Delta_2)l^\mu=-\delta l^\mu=0. \label{Del12l}
	\end{align}
	This extension is tied to the contravariant tangent. The corresponding identity is not preserved in general by lowering the index, although it is preserved when $\xi$ is Killing.
	
	Commutation relations between covariant variations and the worldline covariant derivative $\nabla_\tau$ play a key role in the subsequent analysis.
	To state such identities unambiguously, one must specify how the variations act on every object entering $\nabla_\tau$, in particular on the tangent vector.
	For a spacetime field $\phi$ pulled back to the worldline, one finds
	\begin{equation}
		\Delta_1 \nabla_\tau \phi - \nabla_\tau \Delta_1 \phi = \epsilon\, R^\alpha{}_{\beta\mu\nu}\, \xi^\mu l^\nu G^\beta_\alpha\, \phi ,
		\label{com1}
	\end{equation}
	where the left-hand side is evaluated using the extension $\Delta_1 \al \equiv \Delta_2 \al = \epsilon\, \nabla_\al \bm{\xi}$ [cf.\ Chap.~5 of Ref.~\citenum{DeWitt:Book:2011}]. For a scalar field $a(x)$, the curvature term vanishes, and Eq.~\eqref{com1} reduces to $\delta_1\dot a = d(\delta_1 a)/d\tau$, as expected.
	
	A curvature-type commutation relation analogous to Eq.~\eqref{com1} does not hold for $\Delta_2$ with an arbitrary vector field $\bm{\xi}$.
	For a generic tensorial object $\Phi$,
	\begin{equation}\label{com2}
		\Delta_2 \nabla_\tau \Phi - \nabla_\tau \Delta_2 \Phi = - \epsilon\, l^\lambda \xi^\nu_{;\mu\lambda} G^\mu_\nu\, \Phi.
	\end{equation}
	However, if $\xi^\mu$ is a Killing vector field, its second derivative satisfies the integrability condition
	\begin{equation}\label{eq:KillingId}
		\nabla_\mu \nabla_\nu\, \xi^\rho = R^\rho{}_{\!\nu\mu\sigma}\, \xi^\sigma .
	\end{equation}
	Using the antisymmetry of the Riemann tensor in its last two indices, Eq.~\eqref{com2} then becomes
	\begin{equation}
		\Delta_2 \nabla_\tau \Phi - \nabla_\tau \Delta_2 \Phi
		\stackrel{\xi_{(\mu;\nu)}=0}{=}
		\epsilon\, \tensor{R}{^\nu_\mu_\sigma_\lambda}\,\xi^\sigma l^\lambda G^\mu_\nu\, \Phi .
	\end{equation}
	For an ordinary scalar $A$ (density weight zero), $G^\mu_\nu A=0$ and therefore $\Delta_2 A=0$.
	
	The total variation $\delta_2$, and consequently its covariant extension $\Delta_2$, does not allow independent variations of worldline tensors: their variation is fixed by the diffeomorphism that displaces the worldline and induces the corresponding transformation of the tensors attached to it.
	In many applications, however, worldline objects must be allowed to vary independently of the trajectory, with the reduction to the physical degrees of freedom imposed later through constraints.
	
	To separate the variation induced by moving the worldline from these independent variations, while retaining the desired properties of spacetime variations, we define the extended shift variation $\hat{\delta}_1$ by prescribing its action as
	\begin{align}
		&\hat{\delta}_1 \phi = \delta_1 \phi, \\
		&\hat{\delta}_1 l^\mu = \delta_2 l^\mu, \qquad \hat{\delta}_1 l_\mu = \g_{\mu\nu} \delta_2 l^\nu, \\
		&\hat{\delta}_1 \psi = \delta_\|\psi,
		\label{eq:hat.delta1.psi}
	\end{align}
	on fields, tangent vectors, and all other worldline tensors, respectively. The extended covariant variation is then defined by
	\begin{equation}
		\hat{\Delta} \Phi \defeq \hat{\delta}_1 \Phi - \delta_\|\Phi.
		\label{def:hat.Delta}
	\end{equation}
	For an independent worldline tensor $\psi$, this is equivalent to extending the shift variation \eqref{def:shift.variation} by defining the value at the displaced point through parallel transport, $\psi(\bar{x})\equiv\psi_\|(\bar{x})$, which trivially satisfies $\hat{\Delta}\psi\equiv 0$.
	This statement applies to the independent tensor $\psi$, but not directly to the composite quantity $\nabla_\tau\psi$, which also contains the tangent vector and the connection.
	Using
	\begin{equation}
		\hat{\delta}_1 \psi =
		-\epsilon\,\Gamma^\nu_{\mu\rho}\xi^\rho G^\mu_\nu\psi,
	\end{equation}
	we have
	\begin{equation}
		\hat{\delta}_1 \dot{\psi} = \dot{\psi}(\bar{x}) - \dot{\psi}(x) = - \epsilon\, \frac{d}{d\tau} \big(\Gamma^\nu_{\mu\rho} \xi^\rho G^\mu_\nu\psi\big).
	\end{equation}
	It follows that
	\begin{align}
		\begin{aligned}
			\hat{\delta}_1(\nabla_\tau \psi) &= \hat{\delta}_1 \dot{\psi} + \hat{\delta}_1\big(\Gamma^\nu_{\mu\rho}l^\rho G^\mu_\nu\psi\big) \\
			&= \epsilon\left(\partial_\rho\Gamma^\nu_{\mu\sigma} - \partial_\sigma \Gamma^\nu_{\mu\rho}\right) \xi^\rho l^\sigma G^\mu_\nu \psi \\
			& \qquad + \Gamma^\nu_{\mu\rho}l^\rho G^\mu_\nu\hat{\delta}_1\psi - \epsilon\,\Gamma^\nu_{\mu\rho}\xi^\rho G^\mu_\nu\dot{\psi}.
		\end{aligned}
	\end{align}
	The corresponding extended covariant variation is
	\begin{equation}
		\hat{\Delta}(\nabla_\tau\psi)
		= \hat{\delta}_1(\nabla_\tau \psi) + \epsilon\,\Gamma^\nu_{\mu\rho}\xi^\rho G^\mu_\nu \nabla_\tau \psi.
	\end{equation}
	Using the commutator \eqref{eq: Rep Commutator}, the noncovariant connection terms combine into the Riemann tensor, and we obtain
	\begin{equation}
		\hat{\Delta}(\nabla_\tau\psi)
		= \epsilon\, \tensor{R}{^\alpha_\beta_\mu_\nu}\, \xi^\mu l^\nu G^\beta_\alpha\,\psi.
	\end{equation}
	Since $\nabla_\tau\hat{\Delta}\psi=0$, this gives the same curvature-type commutator as Eq.~\eqref{com1}, now for the extended variation $\hat{\Delta}$ acting on independent worldline tensors.
	
	To complete the variational framework, we allow worldline tensors to vary independently of the trajectory.
	Following Ref.~\cite{Steinhoff:AnnPhys:2011}, we denote this variation by $\delta_x\Phi$, where the subscript indicates that the worldline $x^\mu(\tau)$ is held fixed.
	Since we consider propagation on a fixed background, $\delta_x\, \tensor{\g}{_\mu_\nu}\equiv0$ and therefore $\delta_x\Gamma^\mu_{\nu\rho}\equiv0$, such that
	\begin{equation}
		[\delta_x,\nabla_\tau]=0,
	\end{equation}
	which simplifies the analysis significantly.
	
	For an ordinary scalar function $L$ defined along the worldline, such as a worldline Lagrangian, we define the induced same-point variation analogously to Eq.~\eqref{delta-r} by
	\begin{align}
		\delta L \defeq \delta_2 L - \hat{\delta}_1 L .
	\end{align}
	Since $L$ has density weight zero and parallel transport acts trivially on scalars, $\delta_2L=0$ and $\delta_\|L=0$. Thus,
	\begin{align}
		\delta L = - \hat{\delta}_1 L = - \hat{\Delta} L .
	\end{align}

	\section{Applications of covariant variations} \label{S:app}
	In this section, we demonstrate how appropriately defined covariant variations lead to compact derivations of standard results, including the MPD equations~\cite{Mathisson:ActaPhysPol:1937,Papapetrou:ProcRoySocLondA:1951,Dixon:ProcRoySocLondA:1970,Dixon:PhilTransRoySocLondA:1974}.
	In Sec.~\ref{sec:Lagrangian.mechanics}, we use this formalism to obtain a Lagrangian whose Euler--Lagrange equations exactly reproduce the first-order post-eikonal ray-propagation equations.
	
	\subsection{Spinless particles} \label{subsec:spinless.particles}
	First, we consider the worldline Lagrangian of a point particle,
	\begin{equation}\label{eq: L point particle}
		L_\text{kin} = \half \g_{\mu\nu}(x) l^\mu l^\nu.
	\end{equation}
	For the purposes of this subsection, its mass is irrelevant, and we omit the related constraints.
	We illustrate our formalism in two equivalent ways: first, viewing $L_\text{kin}$ as a scalar, we have
	\begin{align}
		\delta L_\text{kin} &= - \hat{\delta}_1 L_\text{kin} = -\hat{\Delta} L_\text{kin} \\
		&= - \epsilon\, \g_{\mu\nu} l^\rho \nabla_\rho \xi^\mu l^\nu = - \frac{d}{d\tau}(\epsilon\, l_\mu \xi^\mu) + \epsilon\, \xi^\mu \nabla_\tau l_\mu. \nonumber
	\end{align}
	This gives the equation of motion $\nabla_\tau l^\mu=0$ and the presymplectic potential
	\begin{equation}
		i_\xi \Theta = l_\mu \xi^\mu.
	\end{equation}
	
	Alternatively, because $L_\text{kin}$ is a scalar and contains no independent worldline tensors $\psi$, the variation can be written in terms of the standard Lie derivative \eqref{Lie}.
	This form is particularly useful when considering spacetime symmetries acting on the theory:
	\begin{align}
		\begin{aligned}
			\delta L_\text{kin} &= (\Delta_2-\Delta_1) L_\text{kin} = \half (\Delta_2 \g_{\mu\nu}) l^\mu l^\nu \\
			&= - \epsilon\,\xi_{(\mu;\nu)} l^\mu l^\nu = -\frac{d}{d\tau} (\epsilon\, l_\mu \xi^\mu) + \epsilon\, \xi^\mu \nabla_\tau l_\mu.
		\end{aligned}
	\end{align}
	In this form, it is manifest that $\delta L_\text{kin}=0$ when $\bm{\xi}$ is a Killing vector field.
	Noether's theorem then yields the standard conserved charge
	\begin{equation}
		Q_\xi = i_\xi \Theta = l_\mu \xi^\mu .
	\end{equation}
	
	\subsection{Massive spinning particles} \label{subsec:massive.spinning.particles}
	In the pole-dipole approximation, where quadrupole and higher multipole moments are neglected, the motion of a classical massive spinning body represented by a worldline $x^\mu(\tau)$ is governed by the MPD equations~\cite{Mathisson:ActaPhysPol:1937,Papapetrou:ProcRoySocLondA:1951,Dixon:ProcRoySocLondA:1970,Dixon:PhilTransRoySocLondA:1974},
	\begin{align}
		&\nabla_\tau p_\mu = - \half R_{\mu\nu\rho\sigma}\, l^\nu S^{\rho\sigma},
		\label{eq:MPD.p} \\
		&\nabla_\tau S^{\mu\nu} = p^\mu l^\nu - p^\nu l^\mu.
		\label{eq:MPD.S}
	\end{align}
	The explicit form of the momentum $p_\mu$ and the spin tensor $S^{\mu\nu}$ depends on the model.
	
	These ten evolution equations do not, by themselves, determine the fourteen variables $x^\mu$, $p_\mu$, and $S^{\mu\nu}$.
	To select a representative centre-of-mass worldline, one must impose a spin supplementary condition (SSC)~\cite{Steinhoff:Chapter:2015}, typically of the form
	\begin{equation}
		S^{\mu\nu} t_\nu=0,
		\label{eq:SSC}
	\end{equation}
	where the timelike vector $t^\mu$ specifies the observer with respect to whom the centre of mass is defined.
	Common choices include the Tulczyjew--Dixon condition $t^\mu=p^\mu$~\cite{Tulczyjew:ActaPhysPol:1959,Dixon:ProcRoySocLondA:1970}, the Mathisson--Pirani condition $t^\mu=l^\mu$~\cite{Mathisson:ActaPhysPol:1937,Pirani:ActaPhysPol:1956}, and the Ohashi--Kyrian--Semerák class~\cite{Ohashi:PRD:2003,Kyrian.Semerak:MNRAS:2007}, in which $t^\mu$ is parallel transported along the worldline, $\nabla_\tau t^\mu=0$. Together with preservation of the SSC and a choice of parametrization, this closes the system.%
	\footnote{A detailed discussion of the interpretation of the SSC and related issues is provided in Ref.~\cite{Costa.Natario:Chapter:2015}.}%
	\ In the quantum case, different choices are related to different spin operators and different notions of particle localization (see, e.g., Refs.~\cite{Bauke.etal:PRA:2014,Terno:PRA:2014,Celeri.Kiosses.Terno:PRA:2016,DeRosa.Moretti:LettMathPhys:2024} for modern treatments).
	
	The spin degrees of freedom are modelled using an orthonormal frame $\{e_A^\mu\}$ defined along the worldline,
	\begin{equation}
		\g_{\mu\nu}\, e_A{}^\mu e_B{}^\nu = \eta_{AB},
	\end{equation}
	where $\eta_{AB}$ denotes the Minkowski metric.%
	\footnote{Note that this frame is \emph{not} the usual orthonormal frame/tetrad one introduces as an alternative way to represent dynamical gravitational degrees of freedom. Some authors~\cite{Barausse.Racine.Buonanno:PRD:2009,Steinhoff:AnnPhys:2011,Steinhoff:Chapter:2015} introduce two frames to distinguish these notions, but this is not necessary here as we are not concerned with gravitational degrees of freedom.}\
	The transport of the frame is
	\begin{equation}
		\nabla_\tau e_A{}^\mu = \dot{e}_A{}^\mu + \Gamma^{\mu}{}_{\!\nu\rho}\, \dot{x}^\rho e_A{}^\nu = - \Omega^{\mu\nu}e_{A\nu},
		\label{eq:frame.transport}
	\end{equation}
	with the spin frame angular velocity $\Omega_{\mu\nu}$ defined as
	\begin{equation}
		\Omega^{\mu\nu} \defeq \eta^{AB}\, e_A{}^{\mu}\, \nabla_\tau e_{B}{}^\nu,
		\label{def:Omega.munu}
	\end{equation}
	or, equivalently,
	\begin{equation}
		\Omega_{AB} \defeq  e_{A\mu}\, e_{B\nu}\, \Omega^{\mu\nu}= e_B{}^{\mu}\, \nabla_\tau e_{A\mu} .
		\label{def:Omega.AB}
	\end{equation}
	By construction, $\Omega_{\mu\nu}$ and $\Omega_{AB}$ are antisymmetric, as follows from metric compatibility and orthonormality of the frame.
	
	Preservation of the SSC is one of the main restrictions on the allowed form of the spinning-particle Lagrangian.
	We use the construction of Ref.~\cite{Steinhoff:Chapter:2015}, restricted to point particles, which is the covariant version of the curved spacetime generalization of the relativistic top construction of Refs.~\cite{Hanson.Regge:AnnPhys:1974,Porto:PRD:2006}.
	The Lagrangian is taken to depend on the worldline variables through scalar invariants constructed from $l^\mu$, $\Omega^{\mu\nu}$, and the fixed background metric, $L=L(\g_{\mu\nu},l^\mu,\Omega^{\mu\nu})$.
	Restricting to parity-even invariants, we take $L = L(a_1,a_2,a_3,a_4)$, where
	\begin{align}
		\begin{aligned}
			&a_1=l^2, \quad a_2=\Omega^2 =\Omega_{\mu\nu}\Omega^{\mu\nu}, \\
			&a_3=l_\mu \Omega^{\mu\nu}\Omega_{\nu\rho}l^\rho, \quad
			a_4=\Omega_{\mu\nu}\Omega^{\nu\rho}\Omega_{\rho\sigma}\Omega^{\sigma\mu}.
		\end{aligned}
	\end{align}
	
	To vary a Lagrangian in this class while allowing the orthonormal frame $\{e_A^\mu\}$ to undergo independent Lorentz transformations, we use the covariant variation
	\begin{equation}
		\Delta \defeq -(\hat{\Delta}+\delta_x).
		\label{def:covariant.variation}
	\end{equation}
	Here, $\hat{\Delta}$ accounts for the variation induced by displacing the worldline, whereas $\delta_x$ acts on the independent internal degrees of freedom with the worldline held fixed.
	Since the frame is an independent worldline tensor, $\hat{\Delta}e_A{}^\mu=0$.
	In the fixed-background setting $\delta_x \g_{\mu\nu}\equiv0$, we therefore have
	\begin{equation}
		\Delta e_A{}^\mu=-\delta_x e_A{}^\mu.
	\end{equation}
	The independent frame variation preserves the completeness relation,
	\begin{equation}
		\delta_x(e_{A\mu}e^A{}_\nu)=\delta_x\g_{\mu\nu}\equiv0,
	\end{equation}
	and is therefore an infinitesimal Lorentz transformation of the frame.
	We parametrize it by
	\begin{equation}
		\delta\theta^{\mu\nu}\defeq e_A{}^\mu\, \delta_x e^{A\nu}, \qquad
		\delta_x e_A{}^\mu = -\delta\theta^{\mu\nu} e_{A\nu}.
	\end{equation}
	It follows that $\delta\theta^{\mu\nu} \!=\! -\delta\theta^{\nu\mu}$, or, equivalently, $\delta\theta_{AB} \!=\! -\delta\theta_{BA}$, with frame components
	\begin{equation}
		\delta \theta_{AB} \defeq e_{A\mu} e_{B\nu}\, \delta\theta^{\mu\nu} = e_{B\mu}\, \delta_x e_A{}^\mu.
		\label{eq:delta.theta.frame.components}
	\end{equation}
	
	With these conventions, the variation of the Lagrangian is
	\begin{equation}
		\Delta L
		= - p_\mu \hat{\Delta} l^\mu
		- \half S^{\mu\nu}(\hat{\Delta}+\delta_x) \Omega_{\mu\nu},
		\label{eq:Langrangian.variation}
	\end{equation}
	where
	\begin{align}
		p^\mu &\defeq \left.\frac{\partial L}{\partial l_\mu}\right|_\Omega = 2 l^\mu \frac{\partial L}{\partial a_1} + 2 \Omega^{\mu\nu} \Omega_{\nu\rho} l^\rho \frac{\partial L}{\partial a_3},
		\label{def:p} \\
		S^{\mu\nu} &\defeq 2 \left.\frac{\partial L}{\partial \Omega_{\mu\nu}}\right|_\al = 4 \Omega^{\mu\nu} \frac{\partial L}{\partial a_2} + 2 \big( l^\mu \Omega^{\nu\rho} l_\rho - l^\nu \Omega^{\mu\rho} l_\rho \big) \frac{\partial L}{\partial a_3} \nonumber \\
		&\hspace*{35.175mm} + 8 \Omega^{\nu\rho} \Omega_{\rho\sigma} \Omega^{\sigma\mu} \frac{\partial L}{\partial a_4}.
		\label{def:S}
	\end{align}
	We note that if the Lagrangian depends only on $a_1$ and $a_2$, then $p^\mu$ is proportional to $l^\mu$, and no hidden momentum is generated.
	However, to describe generic spinning-particle dynamics compatible with the preservation of an SSC, one must allow invariants coupling $l^\mu$ to $\Omega^{\mu\nu}$, such as $a_3$ \cite{Hanson.Regge:AnnPhys:1974}; in that case the canonical momentum need not be parallel to the kinematic tangent $l^\mu$.

	Using the variational rules introduced in Sec.~\ref{sec:covariant.variation}, we first compute the part of the variation induced by displacing the worldline:
	\begin{align}
		\begin{aligned}
			S^{\mu\nu} \hat{\Delta} \Omega_{\mu\nu} &= S^{\mu\nu} \big[ \hat{\Delta} (e_{A\mu}) \nabla_\tau e^{A}{}_{\nu} + e_{A\mu} \hat{\Delta} \nabla_\tau e^{A}{}_{\nu} \big] \\
			&= -\epsilon\, S^{\mu\nu} e_{A\mu} R^\alpha{}_{\nu\rho\sigma} \xi^\rho l^\sigma  e^A_{~\alpha} \\
			&= -\epsilon\, R_{\mu\nu\rho\sigma}\, \xi^\rho l^\sigma S^{\mu\nu}.
		\end{aligned}
		\label{eq:worldline.variation}
	\end{align}
	The trajectory-independent frame variation gives%
	\footnote{Although the derivation below is written in terms of the frame vectors, the fixed-background result can be expressed entirely through the antisymmetric Lorentz rotation parameter $\delta\theta^{\mu\nu}$.
		If metric variations $\delta_x \g_{\mu\nu} \neq 0$ were included, the same definition $\delta\theta^{\mu\nu}\defeq e_A{}^\mu\delta_x e^{A\nu}$ could still be used, but $\delta\theta^{\mu\nu}$ would no longer be purely antisymmetric.
		Varying the completeness relation $e_A{}^\mu e^{A\nu}=\g^{\mu\nu}$ gives $2\delta\theta^{(\mu\nu)}=\delta_x\g^{\mu\nu}$, so only the antisymmetric part represents an independent Lorentz rotation of the frame.
		In that more general setting, $\delta_x\Omega^{\mu\nu}$ can still be written without retaining explicit variations of the frame vectors, using $\delta\theta^{\mu\nu}$ together with the metric-induced variation of the connection.
		This generalization is not needed in the fixed-background analysis considered here.}
	\begin{align}
		S^{\mu\nu} \delta_x \Omega_{\mu\nu}
		&= S^{\mu\nu} \left[ \delta_x (e_{A\mu}) \nabla_\tau e^{A}{}_{\nu} + e_{A\mu} \delta_x \nabla_\tau e^{A}{}_{\nu} \right] \nonumber \\
		&= \nabla_\tau \left( S^{\mu\nu} e_{A\mu}\, \delta_x e^{A}{}_{\nu} \right)
		+ S^{\mu\nu} e^{A}{}_{\mu} [\delta_x,\nabla_\tau] e_{A\nu} \nonumber \\
		&\quad + \left( S^{\mu\rho} \Omega_{\rho}{}^{\nu} - S^{\nu\rho} \Omega_{\rho}{}^\mu - \nabla_\tau S^{\mu\nu} \right) e_{A\mu}\, \delta_x e^{A}{}_{\nu} \nonumber \\
		&= \nabla_\tau \left( S^{\mu\nu}\delta \theta_{\mu\nu} \right) \label{eq:Omega.variation.frame} \\
		&\qquad + \left( S^{\mu\rho} \Omega_{\rho}{}^{\nu} - S^{\nu\rho} \Omega_{\rho}{}^\mu - \nabla_\tau S^{\mu\nu} \right) \delta\theta_{\mu\nu}. \nonumber
	\end{align}
	In the last equality we used $[\delta_x,\nabla_\tau]=0$, which follows from the assumption of a fixed background. Thus the full variation reads
	\begin{align}
		\begin{aligned}
			{-}(\hat{\Delta} + \delta_x) L =& - \frac{d\Theta}{d\tau} {+ \left( \nabla_\tau p_\mu + \half R_{\mu\nu\rho\sigma}\, l^\nu S^{\rho\sigma} \right) \epsilon\,\xi^\mu} \\
			& - \half \left( S^{\mu\rho} \Omega_{\rho}{}^{\nu} - S^{\nu\rho} \Omega_{\rho}{}^\mu - \nabla_\tau S^{\mu\nu} \right) \delta\theta_{\mu\nu}.
		\end{aligned}
		\label{eq:Omega.variation.full}
	\end{align}
	The total derivative term identifies the presymplectic potential evaluated on this variation as
	\begin{align}
		\Theta = \epsilon\,p_\mu \xi^\mu+\frac{1}{2}S^{\mu\nu}\delta\theta_{\mu\nu}.
		\label{eq:pre.sym.spin}
	\end{align}
	The coefficient of $\epsilon\,\xi^\mu$ gives the momentum equation
	\begin{align}
		\nabla_\tau p_\mu = -\half R_{\mu\nu\rho\sigma}\, l^\nu S^{\rho\sigma},
	\end{align}
	while the coefficient of $\delta\theta_{\mu\nu}$ gives
	\begin{align}
		\nabla_\tau S^{\mu\nu} = S^{\mu\rho}\Omega_{\rho}{}^\nu - S^{\nu\rho}\Omega_{\rho}{}^\mu = 2S^{[\mu}{}_{\rho}\Omega^{|\rho|\nu]}.
	\end{align}
	Using Eqs.~\eqref{def:p} and \eqref{def:S}, the spin equation can be rewritten as
	\begin{align}
		\nabla_\tau S^{\mu\nu}
		&= 2 S^{[\mu}{}_{\rho} \Omega^{|\rho|\nu]} \nonumber \\
		&= 8 \Omega^{[\mu}{}_{\rho} \Omega^{|\rho|\nu]} \frac{\partial L}{\partial a_2}
		+ 16 \Omega_{\rho}{}^{\sigma} \Omega_{\sigma\lambda}
		\Omega^{\lambda[\mu} \Omega^{|\rho|\nu]} \frac{\partial L}{\partial a_4} \nonumber \\
		& \qquad + 4 \big( l^{[\mu} \Omega_{\rho\sigma} l^\sigma - l_\rho \Omega^{[\mu}{}_{\sigma} l^\sigma \big) \Omega^{|\rho|\nu]} \frac{\partial L}{\partial a_3} \nonumber \\
		&= -4\, l^{[\mu}\Omega^{\nu]\rho}\Omega_{\rho\sigma}l^\sigma
		\frac{\partial L}{\partial a_3}
		= p^\mu l^\nu - p^\nu l^\mu.
	\end{align}
	Thus the covariant variation \eqref{def:covariant.variation} recovers the MPD equations \eqref{eq:MPD.p} and \eqref{eq:MPD.S}.
	
	Similarly to the spinless case, application of $\Delta_2$ simplifies identification of the conservation laws via the Noether theorem.
	For a Killing vector $\bm{\xi}$ we require that the Lagrangian is invariant under the induced diffeomorphism,
	\begin{equation}
		\Delta L=(\Delta_2+\Delta)L\equiv 0.
	\end{equation}
	Taking into account Eq.~\eqref{d2g}, we have
	\begin{align}
		\begin{aligned}
			(\Delta_2 +\Delta) L &= p_\mu(\Delta_2 -\hat\Delta) l^\mu \label{varS} + \frac{\partial L}{\partial \g_{\mu\nu}}\Delta_2 \g_{\mu\nu} \\
			& \quad + \frac{1}{2} S^{\mu\nu}(\Delta_2+\Delta) \Omega_{\mu\nu}.
		\end{aligned}
	\end{align}
	Proceeding analogously to Eq.~\eqref{eq:Omega.variation.frame} and taking into account Eq.~\eqref{com2}, one arrives at
	\begin{align}
		S^{\mu\nu}\Delta_2\Omega_{\mu\nu} =&\epsilon\nabla_\tau( S^{\mu\nu}\nabla_{\mu}\xi_\nu )+ \epsilon  S^{\mu\nu} l^\sigma \nabla_\sigma\nabla_\nu \xi_\mu \label{d2Om} \\
		& + \epsilon (S^{\mu\rho} \Omega_{\rho}{}^{\nu} - S^{\nu\rho} \Omega_{\rho}{}^\mu - \nabla_\tau S^{\mu\nu})\nabla_{\mu}\xi_\nu . \nonumber
	\end{align}
	The first term on the right-hand side of Eq.~\eqref{varS} vanishes by Eqs.~\eqref{eq:hat.delta1.psi} and \eqref{def:hat.Delta}, since $\hat{\Delta}l^\mu=\Delta_2l^\mu$. The second term is proportional to $\pounds_\xi\g$ and is zero as $\bm{\xi}$ is a Killing vector. In the expression for $S^{\mu\nu}(\Delta_2+\Delta) \Omega_{\mu\nu}$, the second term of Eq.~\eqref{d2Om} cancels with the right-hand side of Eq.~\eqref{eq:worldline.variation} due to the Killing property \eqref{eq:KillingId}. Finally, to ensure the vanishing of $\Delta L$ one has to identify the parameters of the  Lorentz transformations to be
	\begin{equation}
		\delta \theta_{\mu\nu} = \epsilon\, \nabla_\mu\xi_\nu,\label{theta-xi}
	\end{equation}
	which cancel the remaining terms.
	
	Inserting this relation into the presymplectic potential \eqref{eq:pre.sym.spin}, one obtains the conserved quantity
	\begin{equation}
		Q_\xi = \frac{1}{\epsilon} i_\xi \Theta = p_\mu \xi^\mu + \frac{1}{2}S^{\mu\nu}\nabla_\mu \xi_\nu.
	\end{equation}
	It can, of course, be checked that this is conserved directly from the equations of motion.

	\section{Lagrangian mechanics of post-eikonal light rays}  \label{sec:Lagrangian.mechanics}
	In this section, we address the main goal of this article: deriving a simple worldline Lagrangian whose Euler--Lagrange equations reproduce the first-order post-eikonal ray equation. This equation describes null, generally nongeodesic trajectories associated with high-frequency circularly polarized solutions of the Maxwell equations.  Several equivalent forms of this ray equation exist, depending on the choice of variables and gauge. We use the form derived by Frolov~\cite{Frolov:PRD:2020} as the target equation to be matched.
	
	This equation is formulated using a Newman--Penrose null tetrad $(\al,\an,\am,\bar{\am})$~\cite{Newman.Penrose:JMathPhys:1962,Chandrasekhar:Book:1992}, where $\al$ is tangent to the canonically parametrized null trajectory.
	The auxiliary null vector $\an$ satisfies the normalization condition $\al\cdot\an=-1$, while the complex polarization vectors are orthogonal to both $\al$ and $\an$, and are normalized such that $\am\cdot\bar{\am}=1$. The real and complex polarization bases are related by
	\begin{equation}
		\am = \tfrac{1}{\sqrt{2}} \big(\ave_2+i \ave_3\big),
		\quad
		\bar{\am} = \tfrac{1}{\sqrt{2}} \big(\ave_2-i \ave_3\big),
		\label{eq:m.mbar}
	\end{equation}
	where $\ave_2$ and $\ave_3$, with components denoted by $e^\mu_{(2)}$ and $e^\mu_{(3)}$, are real spacelike polarization vectors defined along the worldline, orthogonal to both $\al$ and $\an$, and normalized by $\ave_i \cdot \ave_j=\delta_{ij}$ for $i,j=2,3$.
	The polarization can therefore be described either by the real pair $(\ave_2,\ave_3)$ or by the complex pair $(\am,\bar{\am})$.
	Together with $\al$ and $\an$, the former gives a real null-adapted tetrad, while the latter gives the equivalent complex Newman--Penrose tetrad.
	
	At order $1/\omega$, the trajectory equation takes the particularly simple form
	\begin{equation}
		w^\mu \defeq \nabla_\tau l^\mu
		= -\tfrac{\sigma}{\omega} R^\mu{}_{\nu\alpha\beta} l^\nu e_{(2)}^\alpha e_{(3)}^\beta
		= -is R^\mu{}_{\nu\alpha\beta} l^\nu m^\alpha \bar{m}^\beta,
		\label{eq:master}
	\end{equation}
	where $\sigma=\pm1$ labels the helicity and $s\defeq\sigma/\omega$.
	The required tetrad propagation conditions can be expressed in terms of the complex acceleration coefficient
	\begin{equation}
		\kappa \defeq -\aw\cdot\am
		= -is R_{\mu\nu\alpha\beta} l^\mu m^\nu m^\alpha \bar{m}^\beta.
		\label{F-prop}
	\end{equation}
	The remaining gauge freedom in the choice of the null tetrad vectors allows us to impose
	\begin{align}
		&\nabla_\tau\an=0, \label{tetrad-prop-n} \\
		&\nabla_\tau \am=-\kappa \; \an , \quad
		\nabla_\tau \bar\am=-\kappa^* \an,
		\label{tetrad-prop-m}
	\end{align}
	along the trajectory.
	In terms of the same coefficient $\kappa$, the trajectory equation can be written as
	\begin{equation}
		\nabla_\tau \al = -(\kappa^* \am + \kappa \;{\bar{\am}}).
	\end{equation}
	
	Unlike the case of massive spinning particles, where an SSC reduces the spin tensor to three independent components, the post-eikonal light ray model considered here encodes polarization through the helicity parameter $\sigma=\pm1$. We enforce the corresponding polarization condition explicitly through a constraint term in the Lagrangian.
	Once such a constraint term is introduced, it is natural to impose the remaining null tetrad gauge and propagation conditions in the same way.
	This allows us to use a simpler Lagrangian than in the generic spinning-particle case, without introducing additional invariants such as $a_3$ and $a_4$.
	A prescription for constructing a frame satisfying these conditions is given in Ref.~\cite{Murk.Terno.Vadapalli:PRD:2025}.
	In the absence of invariants coupling $l^\mu$ to the polarization variables, the momentum associated with the non-constraint part of the model is proportional to $\al$.
	
	In addition, the gauge conditions \eqref{tetrad-prop-n} and \eqref{tetrad-prop-m} imply that the quadratic spin invariant $a_2$ vanishes on shell.
	To derive the desired propagation equation, it is sufficient to retain the part of $a_2$ associated with rotations in the transverse polarization plane, $a_2 \simeq 2\mathfrak{a}^2$,
	\begin{equation}
		\mathfrak{a} \defeq e_{(2)\mu}\nabla_\tau e_{(3)}^\mu
		= \frac{i}{2} \big( m_\mu \nabla_\tau \bar{m}^\mu -\bar{m}_\mu \nabla_\tau m^\mu \big),
		\label{def:angular.velocity}
	\end{equation}
	where $\simeq$ denotes equality modulo the imposed gauge and transport constraints.
	We therefore take the spin-orbit coupling term to be linear in $\mathfrak{a}$.
	To keep the $\hat{\Delta}$ variation of the tetrad compatible with the parallel transport prescription, we introduce an auxiliary vector $\ak$ that is constrained to be null and identified with $\al$ only on shell. Omitting the gauge-enforcing constraint terms for the moment, we take
	\begin{equation}
		L_\text{dyn} = L_\text{kin} + L_\text{spin},
	\end{equation}
	with
	\begin{align}
		& L_\text{kin} = \omega(k_\mu l^\mu - \tfrac{e}{2} k_\mu k^\mu), \\
		& L_\text{spin} = C \mathfrak{a}, \label{lspin}
	\end{align}
	where $e(\tau)$ is a Lagrange multiplier, and $C$ is fixed to $-\sigma$ as we show below.
	
	Varying $L_\text{kin}$ with respect to $e(\tau)$ establishes the null condition $\ak^2=0$, while variation with respect to $k_\mu$ leads to the on shell identification $l^\mu=e k^\mu$. We fix the parametrization by setting $e\equiv1$. Continuing to the spacetime variation, we find
	\begin{align}
		\hat{\Delta} L_\text{kin} &= \omega k_\mu\hat{\Delta} l^\mu = \epsilon\omega k_\mu l^\nu \xi^\mu_{;\nu}, \\
		C \hat{\Delta} \ga &= i\, \epsilon\, C R_{\mu\nu\alpha\beta}\, \xi^\mu l^\nu m^\alpha \bar{m}^\beta .
	\end{align}
	Integrating the first term by parts,
	\begin{align}
		\hat{\Delta} L_\text{kin}
		= \frac{d}{d\tau}(\epsilon\omega k_\mu\xi^\mu)
		- \epsilon\omega\xi^\mu\nabla_\tau k_\mu .
	\end{align}
	Using $\delta L=-\hat{\Delta}L$ for the scalar worldline Lagrangian, the coefficient of $\epsilon\,\xi^\mu$ gives
	\begin{align}
		\omega\nabla_\tau k_\mu - i\, C R_{\mu\nu\alpha\beta}\, l^\nu m^\alpha\bar m^\beta=0.
	\end{align}
	With $\ak=\al$ on shell, this reproduces Eq.~\eqref{eq:master} for $C=-\sigma$.
	
	The trajectory-independent Lorentz frame variation does not introduce any independent spin dynamics.
	After imposing the transport and gauge conditions, the only surviving frame variation is the residual rotation in the transverse $(e_{(2)},e_{(3)})$ plane.
	We parametrize this rotation by
	\begin{equation}
		\theta \defeq e_{(2)\mu} \delta_x e_{(3)}^\mu=-\delta\theta_{23}.
		\label{def:theta}
	\end{equation}
	The corresponding angular velocity
	\begin{equation}
		\ga=e_{(2)\mu}\nabla_\tau e_{(3)}^\mu
	\end{equation}
	then transforms as
	\begin{equation}
		\delta_x \ga=\dot{\theta}.
		\label{eq:delta.x.ga}
	\end{equation}
	Consequently, for $L_\text{spin}=-\sigma\ga$, the variation of the spin part of the action is
	\begin{equation}
		\delta_x S_{\rm spin}=-\sigma\!\int d\tau\,\dot{\theta} = -\sigma\,\theta\Big|_{\tau_i}^{\tau_f}.
	\end{equation}
	Thus the spin contribution changes only by a boundary term and produces no bulk Euler--Lagrange equation. As desired, the evolution of the tetrad is then entirely determined by the imposed transport conditions.
	
	We now identify the spin tensor generated by the helicity term. Defining the transverse bivector
	\begin{equation}
		B^{\mu\nu}\defeq e_{(2)}^\mu e_{(3)}^\nu-e_{(3)}^\mu e_{(2)}^\nu
		=i\left(m^\mu\bar m^\nu-\bar m^\mu m^\nu\right)
	\end{equation}
	and using \eqref{eq:frame.transport}, the angular velocity \eqref{def:angular.velocity} can be written as
	\begin{equation}
		\ga = - \Omega_{\mu\nu}\, e_{(2)}^\mu e_{(3)}^\nu = - \tfrac{1}{2} \Omega_{\mu\nu} B^{\mu\nu}.
	\end{equation}
	Thus the helicity term $L_\text{spin} = C \ga = -\sigma \ga$ gives
	\begin{equation}
		S^{\mu\nu} \stackrel{\eqref{def:S}}{\defeq} 2 \frac{\partial L_\text{spin}}{\partial\Omega_{\mu\nu}} = \sigma B^{\mu\nu} = i \sigma \left(m^\mu\bar m^\nu-\bar m^\mu m^\nu \right).
		\label{eq:spin.tensor}
	\end{equation}
	Consequently, the spin tensor is purely transverse. The normalized bivector satisfies $B^{\mu\nu}B_{\mu\nu}=2=S^{\mu\nu}S_{\mu\nu}$.
	
	Reproducing Eq.~\eqref{eq:master} requires imposing the propagation conditions \eqref{tetrad-prop-n} and \eqref{tetrad-prop-m}.
	This can be achieved by adding constraint terms to the Lagrangian. First, we introduce auxiliary worldline variables $\kappa(\tau)$ and $\bar{\kappa}(\tau)$, treated as independent off shell.
	The constraints for the acceleration components are
	\begin{equation}
		C_\kappa = \kappa + m^\mu \nabla_\tau k_\mu,
		\qquad
		C_{\bar{\kappa}} = \bar{\kappa} + \bar{m}^\mu \nabla_\tau k_\mu .
	\end{equation}
	On the constraint surface, these give
	\begin{equation}
		\kappa = - m^\mu \nabla_\tau k_\mu,
		\qquad
		\bar{\kappa} = - \bar{m}^\mu \nabla_\tau k_\mu .
	\end{equation}
	Since $\bar{m}=m^*$ and $k_\mu$ is real, these relations imply $\bar{\kappa}=\kappa^*$ on shell.
	Given the null tetrad normalization conditions, the condition $\nabla_\tau n=0$ is enforced by
	\begin{align}
		&C_{n1} = k_\mu \nabla_\tau n^\mu, \\
		&C_{n2} = m_\mu \nabla_\tau n^\mu, \\
		&C_{n3} = \bar{m}_\mu \nabla_\tau n^\mu .
	\end{align}
	Finally, we impose
	\begin{equation}
		C_m = \bar{m}_\mu \nabla_\tau m^\mu,
		\qquad
		C_{\bar{m}} = m_\mu \nabla_\tau \bar{m}^\mu .
	\end{equation}
	Together with the preservation of the null tetrad normalization conditions, these constraints leave only the $n^\mu$ component in the decomposition of $\nabla_\tau m^\mu$. The preservation of $\ak\cdot\am=0$, together with $C_\kappa \!=\! 0$, then fixes this component to be $-\kappa$, giving the first relation in Eq.~\eqref{tetrad-prop-m}. The complex-conjugate relation follows from $C_{\bar{\kappa}}=0$.
	
	To prevent the constraint multipliers from contributing to the equations of motion and conservation laws on shell, we impose their vanishing through an auxiliary multiplier $\Lambda$. A convenient real choice for the constraint Lagrangian is
	\begin{align}
		L_\text{constr}
		&= \lambda_{n_1} C_{n1} + \bar{\lambda}_\kappa C_\kappa + \lambda_\kappa C_{\bar{\kappa}} + \bar{\lambda}_n C_{n2} + \lambda_n C_{n3} \nonumber \\
		& \quad + \bar{\lambda}_m C_m + \lambda_m C_{\bar{m}} \\
		& \quad + \Lambda \big( \lambda_{n_1}^2 \!+\! |\lambda_\kappa|^2 \!+\! |\lambda_n|^2 \!+\! |\lambda_m|^2 \big). \nonumber
	\end{align}
	
	\section{Discussion and conclusions} \label{sec:discussion.and.conclusions}
	In this work, we developed a covariant variational prescription for point-particle and worldline actions in curved spacetime [Sec.~\ref{sec:covariant.variation}]. The main conceptual point is that a variation of a tensorial object is not only a displacement of a point on the spacetime manifold, but must include a rule for comparing tensors in different tangent spaces.
	Parallel transport supplies the required comparison map by bringing tensorial data to a common tangent space before the difference is taken. This allows the variational calculus to be organized in terms of covariant objects from the outset and provides a clean separation between the variation of spacetime fields, the variation induced by spacetime symmetries, and the independent variation of internal worldline degrees of freedom. In particular, $\Delta_1$ [Eq.~\eqref{eq:Delta1.explicit}] is the natural covariant shift variation for spacetime fields and keeps the metric inert, while $\Delta_2$ [Eq.~\eqref{eq:Delta2.explicit}] is tied to the total diffeomorphism action and is therefore the appropriate object for Killing symmetries and their associated charges.
	The extended variation $\hat{\Delta}$ [Eq.~\eqref{def:hat.Delta}] is needed for genuine worldline tensors, such as tetrad or polarization vectors, whose variation is not determined by the displacement of the worldline alone.
	
	The broader significance of these results is that they bring several effective descriptions into a common language.
	Geodesic motion, MPD dynamics, and polarization-corrected light propagation can all be treated as worldline variational problems once the relevant covariant variations are chosen carefully. This is useful both conceptually and practically. Conceptually, it clarifies which parts of the dynamics are consequences of covariance, which parts follow from the choice of internal variables, and which parts are imposed by constraints. Practically, it provides a systematic way to construct actions, presymplectic potentials, and conserved quantities for effective ray models that arise from high-frequency expansions of field equations.
	This is especially relevant for gravitational spin Hall effects, where polarization-dependent corrections are small locally but may accumulate over long propagation distances or near compact objects~\cite{Oancea.etal:PRD:2020,Frolov:PRD:2020,Andersson.Oancea:CQG:2023,Murk.Terno.Vadapalli:PRD:2025}.
	
	To illustrate the scope and practical utility of the prescription, we consider two benchmark systems. In the spinless case [Sec.~\ref{subsec:spinless.particles}], the formalism keeps the background geometry, the displacement of the worldline, and the action of spacetime symmetries separate throughout the derivation. For massive spinning particles [Sec.~\ref{subsec:massive.spinning.particles}], it places the usual MPD dynamics \eqref{eq:MPD.p} and \eqref{eq:MPD.S} in a variational framework that makes the geometric origin of both the curvature force and the spin transport equation transparent.
	The corresponding presymplectic potential \eqref{eq:pre.sym.spin} displays the orbital and spin contributions in a form suited to symmetry arguments. In addition to serving as consistency tests, these examples demonstrate how curvature force terms, spin transport, and conserved quantities arise from the same covariant variational structure.
	
	Our main application [Sec.~\ref{sec:Lagrangian.mechanics}] is the construction of a simple local worldline Lagrangian for post-eikonal light rays in the form of Ref.~\cite{Frolov:PRD:2020}. The spin term \eqref{lspin}, with the coefficient fixed to $C=-\sigma$, reproduces the polarization-dependent ray equation \eqref{eq:master} once the null tetrad constraints and propagation conditions are imposed. Physically, the helicity enters through the rotation of the transverse polarization plane, encoded by the angular velocity $\mathfrak{a}$ defined in \eqref{def:angular.velocity}. This term changes only by a boundary contribution under the residual rotation of the transverse basis, but its spacetime variation couples to curvature and produces the polarization-dependent acceleration. The resulting spin tensor \eqref{eq:spin.tensor} is purely transverse and fixed by the helicity.
	The model can therefore be viewed as a constrained massless spinning particle rather than a generic MPD system with independent spin degrees of freedom.
	
	Several directions follow naturally from this work. First, the light ray action can be recast in a more representation independent form to isolate the minimal set of constraints needed to recover equivalent formulations of the spin optics equations beyond the particular Frolov gauge used here. Second, the role of $\Delta_2$ provides a natural route to conservation laws for the post-eikonal model, including both Noether charges from Killing vectors and charges associated with hidden symmetries, such as Killing--Yano and conformal Killing--Yano structures~\cite{Andersson.Gray.Oancea:PRD:2026}. Third, the same formalism can be applied to higher orders in the $1/\omega$ expansion and to other fields, including gravitational and Dirac wave packets~\cite{Andersson.etal:PRD:2021,Oancea.Kumar:PRD:2023}. Such extensions would clarify which features of spin Hall motion are universal consequences of spin-curvature coupling and which depend on the field theory or on the choice of effective worldline variables.
	
	\section*{Acknowledgments}
	SM is supported by the Czech Science Foundation through the JUNIOR STAR grant 25-17250M. The work of FG and DRT is supported by the Schwinger Foundation.
	
	\bibliographystyle{bibstyle}
	\bibliography{covariantvariation-references}
	
\end{document}